# Can AI be Auditable?

Himanshu Verma | *Delft University of Technology, Netherlands* | H.Verma@tudelft.nl

Kirtan Padh | *Technical University of Munich, Germany* | kirtan.padh@tum.de

Eva Thelisson | *AI Transparency Institute, Switzerland* | eva@aitransparencyinstitute.com

## Summary

Auditability is defined as the capacity of AI systems to be independently assessed for compliance with ethical, legal, and technical standards throughout their lifecycle. The chapter explores how auditability is being formalized through emerging regulatory frameworks, such as the EU AI Act, which mandate documentation, risk assessments, and governance structures. It analyzes the diverse challenges facing AI auditability, including technical opacity, inconsistent documentation practices, lack of standardized audit tools and metrics, and conflicting principles within existing responsible AI frameworks. The discussion highlights the need for clear guidelines, harmonized international regulations, and robust socio-technical methodologies to operationalize auditability at scale. The chapter concludes by emphasizing the importance of multi-stakeholder collaboration and auditor empowerment in building an effective AI audit ecosystem. It argues that auditability must be embedded in AI development practices and governance infrastructures to ensure that AI systems are not only functional but also ethically and legally aligned.

## What is Auditability? Why does it matter in AI?

An **audit** is a systematic, independent, and evidence-based evaluation of processes, systems, or entities, typically conducted to assess compliance with predefined standards, regulations, or ethical guidelines. Power (1996) argues that an audit is *"an active process of making things auditable,"* which involves negotiating legitimate processes for an institutionally acceptable knowledge base and creating environments that are receptive to that knowledge base. A broad range of fields (finance, health, education, public policy, etc.) recognize auditing as a legitimate means of demonstrating good management practice (Power, 2003) that can transform management and governance practices from an *"inherently untrustworthy state into a form that auditors and the public can be comfortable with"* (Pentland, 1993). For example, in financial accounting and security, auditing is widely considered to promote trust and transparency (LaBrie and Steinke, 2019). An essential condition for the aforementioned transformation is the social construction of professionalism, auditor independence, and institutional trust in auditing. These conditions are achieved through a collective, collaborative effort involving *"important processual components"* that manifest in extended interactions between auditors and other stakeholders. These interactions are dynamic and constantly evolving through the interplay of *"economic, regulatory, and political pressures for change"* (Pentland, 1993; Power, 2003).

In the context of artificial intelligence (AI), audits aim to assess the ethics, fairness, transparency, accountability, security, and performance of AI systems throughout their lifecycle. For example, ProPublica's audit of the recidivism algorithm used by the U.S. justice system, the Correctional Offender Management Profiling for Alternative Sanctions (COMPAS) system, revealed systemic racial bias within the system against defendants of color (Larson et al., 2016; Dressel and Farid, 2018). Similar incidents have accelerated across different sectors and, when combined with the increasing pervasiveness of AI in various applications and services—including safety-critical systems—have manifested in discourses on the need to audit and ensure the auditability of AI. Furthermore, regulatory initiatives have been undertaken worldwide to address mounting concerns about the potential misuse and harm of AI, particularly in high-risk sectors such as finance, insurance, education, employment, law enforcement, and healthcare. Beyond these industries, the broader societal impact—on democracy, human rights, and the rule of law—has also been a key motivator in shaping various legislative frameworks. For instance, the regulatory efforts concerning AI and AI auditability in the European Union (EU) began with the High-Level Expert Group on Artificial Intelligence (AI HLEG, 2019; 2020), which laid the groundwork for the EU AI Act (2024). Simultaneously, the OECD (Organization for Economic Cooperation and Development) and the Council of Europe worked to establish foundational AI principles, culminating in the creation of the first International Convention on Artificial Intelligence (Council of Europe, 2024a).

Despite numerous and urgent calls for AI auditing, Mökander (2023) argues that it is a relatively new and emergent phenomenon, with the relevant empirical developments taking place in recent years, and that it *"can be defined both in terms of its intended purpose and in terms of its methodological characteristics"*. Functionally, it serves as *"a governance mechanism that can be wielded by different actors for different objectives,"* (also see Brown et al., 2021) and methodologically, it is *"characterized by a structured process whereby an entity's past or present behavior is assessed for consistency with predefined standards, regulations, or norms"* (Mökander and Floridi, 2021; Mökander, 2023).

**Auditability**, on the other hand, refers to the inherent characteristics, underlying processes (technical and organizational), and design features of an AI system that enable or facilitate auditing AI systems. It includes the availability, accessibility, and traceability of the data, decisions, and processes necessary for an effective audit. More specifically, auditability *"enables interested third parties to probe, understand, and review the algorithm's behavior through disclosure of information that enables monitoring, checking, criticism, and identification of risks, biases and errors"* (Li and Goel, 2024). A high level of auditability ensures that AI systems can be audited in a complete and reliable manner, supporting external oversight, fostering trust, and promoting accountability among stakeholders. An alternative conception of auditability is *"reviewability"*, which involves rigorously established processes for technical and organizational documentation and access to system logs that provide contextual information about the AI system and its use to ensure legal compliance (Cobbe et al., 2021). In this chapter, we will consider auditability as an amalgamation of the aforementioned notions of auditability and reviewability.

Although they are often conflated, the notions of auditability, transparency, and explainability are distinct from one another in the context of AI. Transparency refers to the availability of information about an AI system's design, data, decision-making processes, and usage context (Raji et al., 2020; Mökander and Floridi, 2023). Explainability, on the other hand, focuses on making these—technical and decision-making—processes visible and understandable to humans, often in real time (Doshi-Velez and Kim, 2018; Dwivedi et al., 2023). Auditability, by contrast, is the capacity of an AI system, including the processes enabling its development, testing, deployment, and maintenance, to be systematically and independently evaluated for compliance with predefined standards, regardless of whether the system is inherently transparent or explainable. It encompasses the procedural, technical, and organizational conditions—such as documentation, data traceability, access to logs, and model attributes—that make external scrutiny and accountability possible (Mökander and Floridi, 2023; Li and Goel, 2024). Thus, auditability enables formal oversight and remediation when transparency and explainability are lacking. Still, transparency and explainability are critical contributory factors in effective auditability. According to Li and Goel (2024), *"auditability is enhanced through the implementation of traceability, accessibility, reproducibility, transparency, understandability, and explainability in AI systems."*

AI Auditability ensures that stakeholders (AI designers, service providers, operators, procurers) who are involved throughout the AI lifecycle can be held accountable not only for how the AI system is designed or deployed, but also for the manner and intent of its use and outcomes (Leslie et al., 2021). In 2019, the European Commission's High-Level Expert Group on AI (AI HLEG) published its ethics guidelines for trustworthy AI[1] (AI HLEG, 2019). Here, the guideline on accountability, explicitly states that: *"Mechanisms should be put in place to ensure responsibility and accountability for AI systems and their outcomes. Auditability, which enables the assessment of algorithms, data and design processes plays a key role therein, especially in critical applications. Moreover, adequate and accessible redress should be ensured."* Similarly, the AI HLEG (2019) specifies transparency guidelines that encompass AI explainability requirements: *"The data, systems, and AI business models should be transparent,"* and *"AI systems and their decisions should be explained in a manner adapted to the stakeholder concerned. Humans need to be aware that they are interacting with an AI system, and must be informed of the system's capabilities and limitations."* Moreover, the EU AI Act (European Commission, 2024; AI Act, 2024), which came into effect on August 1, 2024, explicitly defines the legal basis for AI auditability within the European Union (EU).

The AI Act establishes key requirements for AI systems to ensure accountability and auditability. These requirements include establishing risk management systems (Article[2] 9), maintaining appropriate governance and data management practices (§ 10), maintaining comprehensive technical documentation and archiving system performance logs (§ 11, 12, and 18), and ensuring transparency by providing users with adequate information (§ 13 and 14). By virtue of its **material scope** (AI Act, 2024; Ch. 1, § 2), the AI Act, which introduces a risk-based regulatory framework, covers different types of AI systems based on the level and type of risk, including prohibited and high-risk AI systems (HRS), general-purpose AI (GPAI) models, and AI systems with transparency requirements. In addition, the **territorial scope** of the AI Act obliges non-EU actors (providers, importers, distributors, manufacturers and users) who wish to place the AI systems in the EU to comply with the AI Act (AI Act, 2024; Ch. 1, § 2). Moreover, the AI Act imposes a conformity assessment requirement for AI systems (§ 47; Thelisson and Verma, 2024), in particular for HRS marketed in the EU. In addition, the HRS must be registered in the EU database and should bear a CE label including a signed declaration of conformity before being placed on the market. Developers of AI systems are also recommended to establish an internal audit function to improve corporate governance and provide assurance of the independence of the organization's board members. This internal audit function is relevant for HRS as well as for GPAI systems. It will make it possible to identify ineffective or inadequate risk management practices. In certain cases related to HRS, a Notified Body (NB) that is appointed and recognized as a legal entity by the EU Member States may perform conformity assessment (by examining technical documentation and the underlying AI development and deployment procedures) to determine if the AI system complies with the requirements set forth in the AI Act (Section 4, § 31; Thelisson and Verma, 2024).

---

[1] These guidelines outline 7 requirements for AI to be considered *trustworthy*. They include, (a) human agency and oversight, (b) technical robustness and safety, (c) privacy and data governance, (d) transparency, (e) diversity, non-discrimination, and fairness, (f) societal and environmental well-being, and (g) accountability.

[2] Hereafter, we will use the **symbol §** to refer to articles of regulatory frameworks and legislation.

The AI Act is not the only regulatory framework that defines and mandates AI audit and auditability requirements. For example, the **U.S. Algorithmic Accountability Act** (US Congress, 2023), enacted during President Biden's term, contains similar provisions for risk and impact assessment of AI systems. However, since then, President Trump signed a new Executive Order titled "Removing Barriers to American Leadership in Artificial Intelligence" on 23 January 2025 (White House, 2025). This Executive Order revokes existing AI policies and directives that are seen as barriers to U.S. AI innovation, economic competitiveness, and national security. AI Auditability is part of it. It also requires the review of policies, directives, and regulations related to the previous Executive Order 14110 from October 2023 (White House, 2023), signed by President Biden, to identify actions that may conflict with the new policy goals. AI Auditability may be one of them. China's **Interim Measures for the Management of Generative AI Services**, or the "AI Measures," took effect on August 15, 2023. They establish centralized, national security-focused regulations for governing generative AI (Migliorini, 2024; White & Case, 2025). The AI Measures include provisions for government-authorized oversight, security assessments, and inspections to ensure the accountability and compliance of AI services with significant public influence. Auditability is ensured through transparent documentation and recordkeeping about AI and datasets. Additionally, AI developers and providers must disclose information about service compliance, including the disclosure of AI-generated content through implicit (i.e., embedded within the metadata) and explicit (i.e., perceptible by users) labeling (White & Case, 2025). Another recent example of AI regulation is Australia's proposed **Mandatory Guardrails for High-Risk AI** and **Voluntary AI Ethics Principles** (Australian Government, 2024; Catania et al., 2024; White & Chase, 2024). The former applies only to HRS and may become legally binding, while the latter is voluntary and applies to all AI systems, including HRS. Despite their differences, both have provisions for transparency, accountability, and auditability. These provisions are established through risk management frameworks, documentation and recordkeeping, data governance, human intervention and oversight, and conformity assessments.

The aforementioned regulatory frameworks are a few examples of enacted and ongoing regulatory efforts with requirements for AI auditability. A comparative analysis of how AI audits and auditability are mandated in various international regulations is beyond the scope of this chapter. For the remainder of this chapter, however, we will use the EU AI Act as a reference. It is already in effect and has an extraterritorial scope. It also has legally binding auditability requirements for a wide range of AI systems, using a risk-based approach.

As mentioned above, the EU AI Act requires AI developers and providers to establish an internal audit. The formal Notified Bodies (NB) serve only as external entities that perform conformity assessments for HRS (AI Act, 2024; § 13, 28-32). Notified Bodies are independent organizations designated by EU member states to assess the compliance of AI products or systems. They are responsible for determining if HRS meet the necessary **conformity assessment** requirements, providing an independent review of the system's compliance. This assessment consists in checking the **technical documentation**, ensuring proper **risk management procedures**, and conducting **testing** and **evaluation**, often including audits of the AI's decision-making process. Notified Bodies are also entitled to perform external audits via conformity assessments in specific cases to ensure compliance with the EU AI Act. Biometric systems (e.g., facial recognition, fingerprint scanning), for example, are considered high-risk and require a conformity assessment by a notified body. Moreover, AI systems used in (a) *critical infrastructure* (e.g., energy, water, and transportation), (b) *law enforcement* (e.g., predictive policing and surveillance), (c) *government services* (e.g., social welfare systems), (d) *healthcare* (e.g., medical devices, diagnostic and decision support systems, surgery robots), (e) *employment* (e.g., systems for hiring, performance evaluation, or promotion decisions), and (f) *education* (e.g., systems for automated grading and admissions) are considered high-risk due to their potential impact on public safety, human rights, and well-being. These systems affect essential services, individual freedoms, health outcomes, career opportunities, and access to education. Therefore, it is essential to ensure their accuracy, fairness, robustness, reliability, and ethics. Notified Bodies are responsible for evaluating these systems, through external audits, to ensure they adhere to strict safety, transparency, non-discrimination, and reliability standards before deployment.

Here, **internal** and **external audits** characterize the different approaches to conducting audits of AI systems. Internal audits are conducted by an organization's own audit team (Raji et al., 2022a; Raji et al., 2022b) and focus primarily on (a) ensuring quality assurance (of source code, training data, and model quality) to improve, iterate, and evaluate risk assessments of AI products being considered for deployment (Raji et al., 2020; Supreme Auditing Institutions, 2023), (b) enabling AI developers to iteratively improve the trustworthiness of AI systems (Cobbe et al., 2021), and (c) assessing compliance with external regulations and mitigating any reputational risks (Sloane et al., 2022; Sloane and Zakrzewski, 2022). Although internal audits are considered important, they are not perceived as credible and independent by external stakeholders and the general public (Sloane et al., 2022; Sloane and Zakrzewski, 2022). This is partly because they can be implemented in different ways in different organizations (Eulerich, 2021), and partly because the results of internal audits are rarely communicated to external stakeholders (Raji et al., 2022a; Raji et al., 2022b).

**External audits**, on the other hand, mitigate the limitations of internal audits, including the subjectivity of AI developers conducting AI audits. External audits are carried out by independent third-party auditors, standards bodies, Notified Bodies (within EU), recognized audit firms, researchers, civil societies, or (non-)governmental organizations (Li et al., 2023; Li and Goel, 2024). These audits result in objective assessment reports (Liu et al., 2022). They focus on the liabilities of the responsible individuals and organizations that develop and/or deploy their AI products and services, incorporate a rigorous and systematic examination of AI pipelines (including datasets, statistical categories, models, technical documentation, system logs, and other relevant information), and may even legally require interviews with various stakeholders to gather evidence (Shneiderman, 2020). Although external AI audits can be conducted prior to the deployment of an AI product or service, the risk of conducting them after AI deployment could result in significant delays in implementing remediation where necessary, and this in turn could result in the propagation of some harmful effects that could not be identified earlier (Liu et al., 2022).

## Why is Auditability challenging in AI? Does it create conflicts with other Responsible AI principles? What makes it so hard to solve?

AI auditing is a relatively new and emerging domain, unlike the established and standardized tools, metrics, processes, and expertise that underlie the auditing of financial, healthcare, or software systems. As a result, there are numerous challenges and barriers to AI auditing and auditability that manifest in four facets of AI development and deployment (Li and Goel, 2024): **technical**, **organizational**, **instrumental** and **methodological**, and **regulatory**.

The **technical challenges** of AI auditing are rooted in the inherent attributes of the AI workflow and are central to its functioning. A significant portion of these challenges are due to the accelerated and continual increase in complexity and opacity of "black box" AI models, including Generative AI models trained on massive datasets. The complexity of these models correlates with the opacity of their underlying mechanisms; in other words, the lack of explainability of black-box models poses significant barriers to AI auditing (Thelisson, 2017; Knowles and Richards, 2021; Herrera-Poyatos et al., 2025; Pégny et al, 2019; Peltovic, 2023). The notion of explainability is a key underlying principle in the EU AI Act, which is considered an essential requirement for the trustworthiness and accountability of AI systems. However, the regulation does not provide specific guidelines, rules, techniques, or enforcement mechanisms for achieving explainability in relation to AI auditing (Pavlidis, 2023). While this leaves ample room for different stakeholders to interpret explainability in relation to audits, it also creates opportunities for prototyping and validating mechanisms that operationalize explainability (Pavlidis, 2023). Besides explainability, another problem with complex models trained on massive datasets is the prioritization of "quantity over quality" of datasets where local and social contexts and nuances are not adequately represented or encoded, creating significant barriers to effective auditing (Bender et al., 2021; Berghoff et al., 2021; Li and Goel, 2024). Although these challenges fall within the realm of the technical conception of AI, they are exacerbated in interaction with organizational processes, which we discuss below.

The **organizational challenges** in auditing AI mainly manifest due to the lack of well-established and transparent corporate governance processes and structures within organizations. Inadequate, incomplete, and substandard AI development and validation lifecycle documentation policies and practices, including a lack of effective system logging mechanisms, contribute significantly to AI auditability challenges (Königstorfer and Thalmann, 2022; Knowles and Richards, 2021; Raji et al., 2020; Li and Goel, 2024). Unlike other highly regulated sectors, such as healthcare, the requirements for creating, maintaining and updating documentation for the AI domain are not clearly specified or rigorously enforced (Li and Goel, 2024; Königstorfer and Thalmann, 2022). Moreover, the practice of documenting the AI lifecycle, especially—training and testing—data creation, data management and processing, and data governance, is considered time-consuming, optional, and unnecessary due to the fast-moving nature of the industry, where developers often work on multiple projects simultaneously (Li and Goel, 2024; Balahur et al., 2022). These poor documentation practices, in turn, lead to several byproducts, including the presence of demographic bias and lack of rigorous validation of datasets by auditors, lack of details about irregularities in data processing and quality assurance, and lack of oversight in general due to poor annotation and labeling practices, malpractice, and inaccuracies (especially in datasets sourced from third-party vendors) (Gebru et al., 2021; Mitchell et al., 2019; Hutchinson et al., 2021). The absence of regulatory frameworks and standardized metrics to assess the quality of datasets and ensure their auditability also contributes to the aforementioned challenges (Chemielinsky et al., 2021). Not only do these organizational decisions pose challenges in isolation, but they also obfuscate the actors and processes that may be responsible and liable for potentially harmful outcomes of AI products and services (Cooper et al., 2022). In some cases, this obfuscation may be intentional to prevent external auditors from examining AI processes, models, and datasets on the grounds of protecting trade secrets (Cooper et al., 2022; Raji et al., 2020). A well-defined and well-regulated collaboration between internal and external auditors with clearly defined roles, including periodic rotation of external auditors to ensure independence and mitigate conflicts of interest, is therefore desirable to mitigate these challenges and also the risk of harmful consequences of AI after deployment (Raji et al., 2020; Li and Goel, 2024).

The AI auditing challenges concerning **instruments and methodologies** are related to the lack of robust and holistic tools, metrics, standards, competencies, and approaches to effectively and rigorously audit AI systems. In a recent interview study with 35 auditors, Ojewale et al. (2024) found that although there are numerous tools available to assist auditors—both internal and external—in the evaluation of AI, these tools are not sufficient to *"[support] the accountability goals of AI auditing in practice"*. In addition, the increasing complexity of AI models has not been adequately matched by the parallel development of advanced audit tools, competencies, and methodologies (Li and Goel, 2024, Ayling and Chapman, 2022). The barriers posed by inadequate and scarce AI audit tools are further reinforced by the lack of comprehensive, harmonized, and standardized metrics (or key performance indicators (KPIs)), quality control frameworks, certifications and educational programs, procedures, and methodologies to equip auditors in evaluating AI processes and systems (Li and Goel, 2024, Ojewale et al., 2024). ISO/IEC[3] 42001 (International Organization for Standardization, 2023) is an example of an international standard that provides guidelines for organizations to systematically establish governance and risk assessment procedures to ensure responsible development and deployment of AI products and services. It requires the AI lifecycle to be assessed for transparency, accountability, privacy, safety, cybersecurity, and identification and mitigation of bias. Moreover, within the framework of the EU AI Act, the European Commission (EC) has mandated the development of harmonized standards to facilitate the conformity and risk assessment as well as the auditing of AI systems (AI Act, 2024; § 40). More specifically, the EC has officially mandated CEN (European Committee for Standardization) and CENELEC (European Electrotechnical Committee for Standardization) to develop the necessary AI standards (AI Act, 2024; European Commission, 2024; Soler Garrido et al., 2024). In their work, CEN and CENELEC are guided by existing international standards, such as ISO/IEC 42001, and incorporate them into their standards development process to ensure alignment and consistency across countries (Future of Life Institute, 2025). Although this work is ongoing at the time of this chapter's writing, future developments in this domain may lower the barriers to AI auditing using standardized tools and methods.

Finally, the **regulatory challenges**, as the name suggests, lie in the realm of AI regulations—both existing and emerging—and the particular conditions, guidelines, and mandates they contain regarding the rigorous assessment of the AI life cycle. First, existing and emerging AI regulations use generic and abstract principles to govern AI and guide its audit by providing a set of requirements for AI transparency, ethics, trustworthiness, and other qualities that are not easily translated into practice by organizations, AI developers, and auditors (Li and Goel, 2024). Second, the underlying principles that guide AI auditing, such as

---

[3] International Organization for Standardization/International Electrotechnical Commission

*openness* (i.e., explainability, transparency) and *universalism* (i.e., fairness, public interest), are in tension with other principles, such as *conservation* (i.e., security, intellectual property) (Yurrita et al., 2022). These tensions create paradoxes that lead to ambiguity for AI developers and auditors and make it challenging to operationalize these high-level principles. Third, the limited frameworks for AI auditing that do exist—due to the instrumental and methodological challenges discussed above—are not aligned in audit scope, metrics, and methodology (Wong et al., 2023). Fourth, existing AI regulation, in particular the EU AI Act, does not mandate independent and external AI audits for all AI systems. External audits are only mandatory for certain high-risk systems (e.g., biometric categorization systems), while for some other AI systems external audits are voluntary and only a compliance and risk assessment should suffice (Scannel et al., 2024; AI Act, 2024; European Commission, 2024; Knowles and Richards, 2021). Fifth, the aforementioned lack of harmonized requirements, combined with the lack of a well-established, globally accepted conceptualization of risk levels, contributes to the ambiguity surrounding AI auditing (Ellul et al., 2021; Berghoff et al., 2021). Finally, there is a lack of regulations that provide adequate protection mechanisms for independent auditors against litigation, particularly with respect to access to information about the AI development cycle (Li and Goel, 2024; Raji et al., 2020). In addition, essential conditions for independent AI audits, such as ensuring auditor independence, disclosure of conflicts of interest, fair selection and rotation of audit firms, are also lacking (Li and Goel, 2024; Ojewale et al., 2024).

In addition to the challenges inherent to auditing most AI products and systems, self-learning AI systems present further challenges. These are the systems that typically use (deep) reinforcement learning (Li, 2017) and continuously improve their performance and adjust their output in response to new and emerging data and interactions with end users. A notable example of a self-learning AI system that learned and reproduced biased, offensive, and toxic behavior after interacting with unfiltered data and malicious users (trolls) is Microsoft's Twitter bot Tay. The Tay chatbot was released in 2016 and began generating racist, sexist, and anti-Semitic tweets within 24 hours of its deployment, and was shut down shortly thereafter (Wolf et al., 2017). These systems are not trained on well-defined and curated datasets that can be audited for the presence of bias and other irregularities. As a result, they are highly susceptible to the effects of corrupted and biased data that may contain harmful elements such as hate speech, extremist views, conspiracy theories, and misinformation (Palmer, 2025). Consequently, these systems are not *"bound by the same guardrails that prevent models"* from learning and reproducing harmful behaviors and cannot operate in an ethical and safe manner (Palmer, 2025). Furthermore, self-learning AI systems suffer from model drift, which leads to degraded performance, and are also vulnerable to adversarial attacks (Gheibi and Weynes, 2024). These attributes of self-learning AI make them particularly problematic to audit because there are no clear and harmonized guidelines, tools, metrics, and regulations to assess their outcomes and performance, and more importantly, they are more likely to cause harmful effects before they can be rigorously audited.

Also, Large Language Models (LLMs), particularly proprietary ones, will introduce a unique auditability challenge. The inability to access training data, coupled with provider-imposed access restrictions, makes it difficult for external auditors to evaluate these systems comprehensively. Without deliberate design choices that support auditability—such as documentation, Model Cards (Mitchell et al., 2019), or audit APIs—the risks of undetected harm and biased behavior will persist, especially as LLMs become embedded in decision-making workflows across domains.

## How is auditability applied in practice? What methods or tools are used?

Before we could examine the application of AI auditability in practice, we need to add a bit more nuance to the different conceptualizations and categorizations of AI audits. The *first* distinction is between a *narrow* and a *broad* conception of AI audits (Mökander, 2023). The **narrow** conception of AI auditing is *"impact-oriented and [focuses] on probing and assessing the outputs of AI systems for different input data"* and is *"well suited for gathering evidence about unlawful discrimination and tend to be underpinned by experimental designs"* (Mökander, 2023). On the other hand, the **broad** conception of AI auditing examines the established corporate governance structures underlying the AI development processes and workflows; in particular, it is *"process-oriented"* and *"[focuses less on the] properties of the AI systems [than] on the governance structures of the organizations that design and deploy them"* (Mökander, 2023). The broad conception of AI audits is considered to be both effective and complete, as it not only enables the detection of erroneous, harmful, and unethical AI outcomes, but also reveals the systemic underpinnings of such AI behaviors. The EU AI Act, for example, encapsulates and mandates the broad conception of AI auditing, in particular by requiring AI developers and providers to establish governance and data management practices (§ 10) and comprehensive technical documentation and archiving of system logs (§ 11, 12, 18). Furthermore, Mökander et al. (2024) argue that although these two conceptions of AI auditing have different focuses and affordances, they are by no means contradictory and can be applied in a complementary and mutually reinforcing manner. In other words, while the broad conception requires collaboration with the organizations developing AI, the narrow conception allows auditors to independently scrutinize the behavior of an AI system without approval or collaboration with the AI providers (Mökander, 2023).

Next, AI audits can be further classified based on the scope of the audit, which defines the baseline against which AI systems are evaluated, and these can be broadly categorized into *ethical*, *legal*, and *technical* AI audits (Mökander, 2023; Mökander and Floridi, 2022). **Ethical (or ethics-based) AI audits** use ethical principles—such as those specified by the AI HLEG (2019) and OECD (2024)—as a "normative baseline" to assess whether or not the AI system conforms to them (Mökander and Floridi, 2022; Mökander, 2023; Li and Goel, 2024). They can be conducted either as a collaborative effort between auditors and AI developers/providers (Raji et al., 2020), or they can be conducted in an adversarial manner, with independent external auditors reviewing the AI system without access to the underlying data, model parameters, and documentation (Jaiswal et al., 2022). Regardless of the manner, they are generally conducted prior to the deployment of AI systems (Li and Goel, 2024). **Legal AI audits** use relevant national and international regulatory frameworks as a baseline to assess whether AI systems and their underlying development processes are in accordance with these frameworks (Mökander, 2023). They may choose to focus on the legal provisions outlined in a single regulation that addresses a specific aspect of the AI lifecycle. For example, this could include data governance and privacy (GDPR, 2016; Data Protection Act, 2018), discrimination (Civil Rights Act, 1964; Employment Equality Directive, 2000; Accessibility Act, 2019), misinformation and content moderation (Digital Services Act, 2022), or cybersecurity (Cybersecurity and Resilience Bill, 2024). Alternatively, they could focus on the entire AI development and deployment process (AI Act, 2024). The **technical AI audits** evaluate the technical qualities of AI systems by comparing them to established benchmarks for accuracy, robustness, safety, explainability, and reliability (Mökander, 2023). These audits can be carried out *ex-ante* (i.e., prior to the deployment of an AI system) or *ex-post* (i.e., continuous monitoring after the deployment of an AI system and the changes in its performance resulting from interactions with new data and users). The EU AI Act, for example, imposes ex-ante conformity and risk assessment requirements, including CE marking for high-risk systems before they are placed on the market, which assess AI systems against ethical, legal and technical principles and benchmarks; at the same time, the AI Act establishes ex-post obligations, requiring continuous monitoring and market surveillance in the event of harmful incidents or malfunctions (Castets-Renard and Besse, 2022; Thelisson and Verma, 2024). It is important to note that these three types of AI audits should not be interpreted as mutually exclusive; rather, they are complementary and can be performed within a single audit. Recent AI regulations, particularly the AI Act (2024), include requirements for all three of these audits.

In addition to the aforementioned audit types, Li and Goel (2024) propose two additional categorizations of *empirical* and *data audits* in their systematic review of AI audits and auditability. **Empirical AI audits** are closer in conception to the adversarial approach of AI audits described above, where AI systems can be evaluated after their deployment by examining the correlations between input and output data that are the consequence of the AI model or algorithm. ProPublica's audit of the recidivism algorithm COMPAS, used by the U.S. justice system, which revealed systemic racial bias within the system against defendants of color (Larson et al., 2016; Dressel and Farid, 2018), is an example of an empirical audit. Moreover, **data audits**, while not strictly within the scope of AI development workflows, still involve the rigorous review of data collection, curation, and processing procedures, as well as the overall assessment of data governance policies within organizations and the responsibilities of data collectors and providers. Data audit requirements are specified in regulations such as the GDPR (2016) and the Data Protection Act (2018), which require data collectors and processors to ensure privacy, security, representativeness, and inclusion, and respect for data owners' consent and rights. These audits use approaches such as Data Protection Impact Assessments (DPIAs), Human Research Ethics Committee (HREC) reviews, and data governance audits to ensure that the data used to train AI models is collected and used fairly and responsibly (Thelisson, 2019; Gebru et al., 2021; Cai et al., 2022; Chmielinski et al., 2022). Recent discourses on AI auditing advocate that ethics-based AI audits should be expanded to include broader **environmental, social, and governance (ESG) audits** (through better corporate social governance and sustainability practices) in AI lifecycle assessment (Minkkinen et al., 2022; Thelisson et al., 2023). This entails not only assessing the traditional AI development workflow, but also incorporating the broader impacts of the deployment and maintenance of AI systems (especially Generative AI) on environmental and societal well-being, in particular as a result of rare earth mineral extraction for chip manufacturing, water and energy use, and carbon footprints.

The above categorization of AI audits based on their scope and conceptualization is a useful tool for clarifying the nuances underlying the complex construct of AI audits. However, it is by no means an exercise in delineating clear boundaries between them. *"The three audit types [i.e., ethical, legal, and technical] are thus best viewed as a continuum of complementary approaches with different focal points"* (Mökander, 2023). The ethical guidelines for trustworthy AI (AI HLEG, 2019) have consolidated these three dimensions, requiring AI systems to be lawful, ethical, and (technically and environmentally) robust. These requirements are widely adopted in various AI regulations (e.g., EU AI Act) and proposed standards (e.g., ISO 42001) to ensure a comprehensive assessment of various facets of AI development, deployment, and operation. Mökander (2023) further argues that this categorization provides different stakeholders with an essential vocabulary of what should constitute an AI audit, in addition to illustrating its main purposes.

Various tools and methodologies have been developed to audit AI systems in one or all of the above audit dimensions. Although it is beyond the scope of this chapter to list them all, we will discuss the most prominent tools and methodologies for AI audits. The European Commission's High-Level Expert Group on AI has compiled a comprehensive checklist of ethical guidelines for trustworthy AI (AI HLEG, 2019) to enable AI developers and service providers to self-assess their systems and ensure compliance with the EU AI Act. The Assessment List for Trustworthy AI[4] (ALTAI) (AI HLEG, 2020) not only consolidates the ethical, legal, and technical dimensions of AI audits, but also includes provisions for data and ESG (Environmental, Societal, and Governance) audits, i.e., ensuring that AI systems are designed to respect environmental and societal goals, policies, and guidelines. ALTAI is intended for AI developers and providers to facilitate internal audits, as well as for external, independent auditors. The AI Transparency Institute[5], a nonprofit organization based in Switzerland, has developed several web-based assessment tools under the name careAI. These tools provide AI developers, service providers, and other relevant stakeholders (including auditors) with the means to translate high-level principles for ethical and trustworthy AI (e.g., AI HLEG, 2019) into quantified metrics and scores to enable development on dimensions that need further attention (Thelisson and Verma, 2024). Their web applications encapsulate various regulatory frameworks (e.g., GDPR, 2016; AI Act, 2024), standards (e.g., ISO/IEC 42001 (ISO, 2023); IEEE 7000 Series (IEEE SA, 2021)), and guiding principles for AI (e.g., ethical guidelines for trustworthy AI (AI HLEG, 2019); EU Charter of Fundamental Rights (FRA, 2009); OECD AI Principles (OECD, 2019)) to enable the assessment of different attributes of AI systems and facets of the AI lifecycle, including corporate governance, corporate digital responsibility, sustainability and environmental impacts of AI, and compliance with EU AI Act (Thelisson et al., 2023; Thelisson and Verma, 2024). Another example of a methodology developed to audit AI in line with the EU AI Act is capAI (Floridi et al., 2022; Mökander and Floridi, 2023). It is primarily aimed at ethical AI audits and proposes a comprehensive and standardized process for auditing the different stages of AI development and deployment by enabling *"the correction of unethical behavior of AI systems and [informing] ethical deliberation throughout the [AI lifecycle]"* (Floridi et al., 2022). An auditability checklist for auditors has also been proposed, specifying the different types of documentation auditors need to assess the different stages of AI development and monitoring (Supreme Auditing Institutions, 2023). Moreover, Constantinides et al. (2024) developed 22 "RAI Guidelines" and a tool for AI developers and practitioners. The goal is to simplify, map, and consolidate the broad range of legal provisions from the EU AI

---

[4] The Assessment List for Trustworthy AI provides a checklist for AI developers and service providers to assess their AI systems across the seven key dimensions – (1) human agency and oversight, (2) technical robustness and safety, (3) privacy and data governance, (4) transparency, (5) diversity, non-discrimination and fairness, (6) environmental and societal well-being, and (7) accountability.
[5] https://www.aitransparencyinstitute.com

Act and different ISO standards to facilitate rapid, usable internal auditing and the development of responsible AI.

In addition to the approaches above that aim to audit the entire AI lifecycle across multiple dimensions, numerous tools and methodologies have also been developed that audit only a specific technical aspect of the AI system, such as training and model development. Examples of instruments designed to assess and report the quality of training datasets, including guidelines for developing, processing, and distributing datasets in a responsible manner, include Datasheets for Datasets (Gebru et al., 2021), Dataset Nutrition Labels (Holland et al., 2020; Chmielinski et al., 2022), and Data Statements (Bender and Friedman, 2018). Similarly, methodologies that enable AI developers and service providers to transparently report key information about their AI models, such as model and performance characteristics, benchmarking results, intended use and context, and conformity declarations, are presented in Model Cards (Mitchell et al., 2019) and Factsheets (Arnold et al., 2019).

# What's an example of auditability in AI done well? What's an example of auditability in AI gone wrong?

Auditable AI encompasses more than the principles of responsible AI (robustness, explainability, ethics, and efficiency); it also provides the documentation and records necessary to withstand regulatory review, as described in the previous sections. There are various examples of positive outcomes resulting from auditability and negative outcomes resulting from a lack of auditability in both the private and public sectors.

### Auditability Gone Wrong

One of the most well-known examples here is COMPAS (Correctional Offender Management Profiling for Alternative Sanctions). Developed by a private company (Northpointe, now Equivant[6]), COMPAS is a proprietary black-box. Its formula and weighting of factors are kept secret, with no disclosure of how it calculates risk (ProPublica, 2016). This means defendants and independent experts cannot fully audit or challenge the basis of a COMPAS score. Only limited documentation (the input questionnaire) is available, and there's no built-in explanation for a given score. The lack of transparency contributed to a major controversy. In 2016, an empirical audit by journalists analyzing COMPAS outcomes found that the tool over-predicted risk for Black defendants (labeling them high-risk at roughly twice the rate of whites who did not reoffend). Meanwhile, white defendants were more often mistakenly labeled low-risk. Northpointe disputed claims of bias, but public debate raged about its fairness because the algorithm was not auditable (Washington, 2018). The Wisconsin Supreme Court allowed COMPAS's use but warned that it should not be the sole basis for decisions and that it must come with explanations of its limitations. Overall, COMPAS illustrates auditability gone wrong: a high-stakes AI deployed without explanation or third-party access, resulting in biased outcomes. There are also multiple examples of opaque predictive policing tools that have had to be scaled back or shut down after years of activism uncovered crucial details about predictive policing. According to the Brennan Center, "*Some of the skepticism around predictive policing programs has less to do with specific technologies than with the lack of transparency from the agencies that administer them—both in terms of what kinds of data are analyzed and how the departments use the predictions*" (Lau, 2020). In other words, the lack of auditability.

---

[6] https://www.equivant.com

Governments have used AI to detect fraud in social services. In the Netherlands, for instance, both local and national agencies deployed machine-learning models to flag citizens who might be committing welfare or benefits fraud (Heikkilä, 2022). The intention was to efficiently target fraud investigations and save public funds. These systems were notoriously non-transparent, however. Citizens did not know an algorithm was scoring them, nor could they access their own "risk score." External audits later revealed that factors like immigration status, single parenthood, or personal debt could unjustly trigger investigations (Burgess et al., 2023). The consequences were severe. In Rotterdam, biased outcomes led to the unjust investigation of many vulnerable individuals, prompting the city to suspend the system. Nationally, the childcare benefits algorithm falsely accused over 20,000 families of fraud, causing widespread financial devastation and ultimately leading to the resignation of the Dutch government in 2021 (Schaart, 2021).

In 2019, Buenos Aires linked its CCTV cameras to a "Fugitive Facial-Recognition System," matching passers-by against a national criminal database. Civil litigation revealed severe misuse: police had improperly queried nearly 15,000 non-fugitives, causing at least 140 wrongful stops or arrests. One victim was jailed for six days due to a clerical error. Lacking oversight, logging, or accuracy records, the system was frozen in 2020 and declared unconstitutional in 2022, a decision upheld on appeal in 2023, pending the implementation of transparency and oversight (Naundorf, 2023).

Kenya faced similar issues with its digital-ID rollout, Huduma Namba. In 2021, the High Court halted the scheme because the government failed to conduct the legally required Data Protection Impact Assessment (DPIA). Despite a 2023 relaunch as Maisha Namba, the program continues to be legally challenged amid concerns about exclusion risks and surveillance potential. Advocates continue to demand comprehensive DPIA and transparency measures, underscoring how bypassing audit processes only delays addressing fundamental governance issues (Macdonald, 2025).

## Auditability Done Right

It is harder to find examples of auditability done right, because often when a thorough audit is performed pre-deployment, the issues are corrected before they become public. Since most AI audits are internal, these examples are not always known publicly. However, there are a few examples where the auditability of a system was improved in response to public concerns, eventually leading to a positive outcome.

Twitter used a saliency-based AI to auto-crop image previews (choosing which part of an image to show in timeline thumbnails). It was meant to focus on the "most interesting" part of photos. Initial audits were informal—in 2020, users noticed the crop often favored showing white faces over Black faces, sparking public experiments. Twitter responded with a commendable transparency move: it open-sourced the algorithm and launched an "algorithmic bias bounty" contest, inviting third parties to audit the code for biases (Field, 2021). All these actions (making code and data available) enhanced auditability for outsiders. The outcome was that both independent researchers and Twitter's in-house team confirmed racial bias in the cropping AI—it was significantly more likely to highlight white faces in images. Twitter publicly acknowledged the issue and stopped using the algorithm for most image previews (introducing manual full-image previews instead). The bias bounty approach was praised as a *positive result*: by embracing auditability and external scrutiny, Twitter not only found and fixed a problem, but also set an example of transparency in social media AI. Another example of an external audit leading to positive change is the Gender Shades study (Buolamwini and Gebru, 2018). Commercial facial analysis APIs (from companies like IBM, Microsoft, and Face++) used AI to classify gender from images—intended for security or photo apps. Two AI researchers, Joy Buolamwini and Timnit Gebru, conducted an independent audit by testing these systems on a diverse set of faces. They examined error rates by demographic group, essentially creating documentation of performance disparities. The study uncovered severe accuracy biases. For instance, IBM's system misidentified the gender of dark-skinned women ~35% of the time versus only 1% for light-skinned men. Exposing this gap pressured companies to act. IBM, for instance, released new diverse training datasets to reduce bias and improved its algorithms (Vincent, 2018). Microsoft and others also reported upgrading their models. Auditability (via third-party scrutiny) brought hidden bias to light and led to fairer AI systems.

Another example is the loan granting process of the Dutch Executive Agency for Education (DUO). Between 2012 and 2023, the Dutch Executive Agency for Education (DUO) employed a simple rule-based risk profiling algorithm to support its student grant verification process. Although designed around three seemingly neutral categorical factors (type of education, student age, and residential distance from parents), the algorithm inadvertently led to indirect discrimination, particularly affecting students with a non-European migration background (Holstege et al., 2025). Notably, concerns about the fairness of the system prompted external audits conducted in collaboration with DUO, which provided access to the necessary data and documentation. These audits revealed the discriminatory impact of the algorithm (Algorithm Audit, 2024). DUO acknowledged the issue, and the Dutch Minister for Education, Culture and Science apologized on behalf of the Dutch government in 2024. A compensation plan of EUR 61 million was also announced, offering refunds to over 10,000 affected students (NL Times, 2024).

The Brazilian national space-research agency (INPE) has open-sourced both the code and geospatial output for its forest monitoring systems. Journalists, NGOs and prosecutors use those open feeds to audit official claims, correlate alerts with illegal-logging raids, and even check the government's own enforcement statistics, creating a well-functioning, crowd-sourced audit loop that has informed enforcement actions since 2004 (TerraBrasilis, *n.d.*).

The above examples demonstrate the importance of AI audits in preventing harm caused by biased outcomes or a lack of transparency in development. While issues can be addressed post-deployment, proactive auditing is far more effective. Positive examples—such as Twitter's open audit of its image-cropping algorithm and the independent Gender Shades study revealing bias in facial recognition—demonstrate how auditability can expose and correct systemic flaws. Even in public institutions, as with DUO's student grant algorithm, external audits enabled by data access can drive accountability, leading to recourse for the affected parties.

## What's missing to make auditability work in AI? What needs to change?

As we previously discussed, AI auditability is a rapidly emerging field. It is evolving in tandem with parallel developments in AI governance principles, regulatory frameworks and standards, and, crucially, the emergence of robust and validated tools and methodologies for auditing AI. Despite the accelerated growth in these areas in recent years, a synchronized alignment among them is still desired, and the practice of AI auditability has a lot of ground to make up (Li and Goel, 2025). In addition, Gabriel and Keeling (2025) advocate for a fair process-based model of AI alignment that emphasizes public standards of justification, context-sensitive AI principles that accommodate diverse moral beliefs, and explanations of AI outcomes and processes that are consistent with human intuition. These proposals respond to the normative challenge of AI alignment (i.e., aligning AI systems with human values and goals to prevent harm), where existing approaches, such as aligning AI with human intentions and values or ensuring that it is helpful, honest, and harmless, have been criticized as incomplete and insufficiently justified, especially in pluralistic societies (Gabriel and Keeling, 2025). Below, we discuss some of these persistent gaps and ongoing efforts—between policy-makers, AI developers, and auditors—to achieve better AI alignment that require further attention in a concerted multi-stakeholder collaboration to enable effective, resilient, and standardized AI auditability.

International regulatory efforts and developments on AI governance and auditability—some of which have been enacted, while others are still in development—do not align on a harmonized and standardized methodology for auditing AI products and systems. While these regulatory frameworks agree on the need for accountability and auditability of AI, the guidelines for the nature and conduct of the audit vary significantly. For example, in the EU AI Act, there is a provision for internal audits requiring conformity and risk impact assessment, with only certain classes of high-risk systems requiring external audits by Notified Bodies before being placed on the market (Thelisson and Verma, 2024). This implies that most, if not all, AI systems will be self-certified by the organizations developing them, with little evidence of how rigorously and transparently these internal audits have been conducted. Although the Notified Bodies can conduct independent external audits, they are still limited in the sense that they can only assess whether the AI systems meet the necessary conformity assessment requirements laid down by the EU AI Act through the evaluation of the established technical documentation, risk management procedures and testing of the AI system's decisions (AI Act, 2024; § 13, 28-32). To address this gap, it is essential that policymakers and standards bodies cooperate globally to align AI regulations. This effort should be supported by AI researchers, developers, and auditors. One example of such collaboration is an international treaty, the Council of Europe's **Framework Convention on AI and Human Rights, Democracy and the Rule of Law**, which was adopted on May 17, 2024 by 46 member states, the European Union, and 11 non-member states, with representatives from the private sector, civil society, and academia as observers (Council of Europe, 2024a). Similarly, parallel efforts focused on consensus building and regulatory alignment to balance economic, social, and environmental goals in AI governance and promote trustworthy and fair AI are taking place within the United Nations (UN), notably the Global Digital Compact[7] (UN, 2024) and AI Governance Day (ITU, 2024). Despite these initial agreements, the international consensus on AI regulation and accountability is volatile and susceptible to geopolitical shifts that undermine established regulatory frameworks and auditability procedures, as evidenced by the recent divergence in AI policy between the US and the EU (Delcker, 2025).

Nevertheless, there are significant ongoing efforts, notably led by the *Ad Hoc* Committee on AI (CAHAI) and the UK's national institute for data science and AI, the Alan Turing Institute[8], to define and recognize a harmonized and structured methodology and tool to guide risk and impact assessment of AI systems (Council of Europe, 2024b; CAI, 2024). This methodology, referred to as the **HUDERIA**, is based on the need to identify, assess, prevent and mitigate risks related to human rights, democracy and the rule of law in the context of AI systems and was adopted by the Committee on Artificial Intelligence (CAI) of the Council of Europe on 28 November 2024 (CAI, 2024). In addition, HUDERIA promotes the alignment between existing and future guidance on standards and international regulatory frameworks such as ISO, IEC, CEN, CENELEC, IEEE, OECD, NIST AI Risk Management Framework (NIST, 2023), and the EU AI Act's Fundamental Rights Impact Assessment (FRIA). Drawing on socio-technical principles, the HUDERIA methodology illustrates a *four-step* approach that includes (a) iterative context-based risk analysis, (b) extended stakeholder engagement, (c) risk and impact assessment on human rights, democracy and the rule of law, and the (d) definition of a mitigation and remediation plan (CAI, 2024). However, signatories to the Framework Convention (Council of Europe, 2024a) may use or adapt it to develop new approaches, provided that they comply with their obligations under the Convention. Similar to this approach, which focuses on human rights, democracy, and the rule of law, there have also been calls to develop frameworks and methodologies that allow for the assessment of AI's impact on the environment and sustainability (Minkkinen et al., 2021; Thelisson et al., 2023). Luccioni et al. (2025) argue that *"[AI] model audits that aim to evaluate model performance and disparate impacts mostly fail to engage with the environmental ramifications of AI models,"* and advocate for the urgent need to expand existing AI regulations by consolidating an audit of the impact of AI systems on planetary resources, carbon footprint, energy consumption, and their disproportionate impact on vulnerable communities and species.

Developments in harmonized regulatory frameworks and technical standards for AI auditability only illustrate a partial and incomplete state of operationalization of AI auditability at scale. The perceptions and impacts of these frameworks and standards on organizations developing and marketing their AI models and products remain largely unexplored. Kilian et al. (2025) conducted an interview study with 23 European start-ups and SMEs to investigate how these organizations are affected by the planned standardization efforts, at least within the EU and within the scope of the EU AI Act. Their findings reveal several challenges faced by these organizations, most notably the disproportionate impact they are perceived to have on startups, particularly in terms of additional barriers to market entry. Furthermore, the standards require a shorter-than-usual timeframe for implementation in terms of compliance, which places an additional burden on SMEs, on top of the high financial costs of complying with the standards. The author's analysis suggests that these challenges may arise from the lack of representation of start-ups and SMEs in standardization bodies and the consequent lack of influence on standardization outcomes. Therefore, policies need to be put in place to ensure (a) increased participation, including incentives, in standardization activities; (b) free and early access to draft harmonized standards, including access to implementation toolkits and evaluation frameworks; (c) subsidies for start-ups and SMEs to support their compliance with the EU AI Act; (d) promotion of goal-based rather than rule- or process-based standards; and (e) better alignment of existing industry-specific (sectoral) standards with AI standards (Kilian et al., 2025).

---

[7] https://www.un.org/global-digital-compact/en
[8] https://www.turing.ac.uk

General Purpose AI[9] (GPAI) systems are considered different from traditional AI products and services in terms of auditability and audit practices. These systems have raised serious concerns in the regulatory and academic communities, primarily because of the serious risks they pose to humanity (e.g., through widespread automation leading to massive layoffs, or their misuse against democratic institutions), and as a result calls have been made to pause, if not halt, their development and critically devise appropriate regulatory safeguards (Future of Life Institute, 2023). Despite these concerns, existing regulations lack concrete guidelines, tools, and methodologies for auditing such models, and the provision of an internal audit will not be sufficient, partly because they are not independent, and partly because GPAI models may adjust (or degrade) their performance to give a false impression of their ability to meet the assessment requirements (Thelisson, 2023). The EU AI Act, for instance, contains obligations for the assessment and auditing of GPAI systems (AI Act, 2024; § 50-55), but these will be enforced one year after the AI Act enters into force (i.e. from August 2025), but the operationalization of GPAI auditing may take even longer because the standardization process (mandated to CEN/CENELEC) will require even more time. In the meantime, § 56 (AI Act, 2024) establishes a set of guidelines, the **AI Act Codes of Practice**, to ensure compliance with the regulation, especially in the interim period between the enactment of GPAI obligations and the implementation of standards. These codes are not legally binding and contain three guidelines for GPAI developers and providers: (1) *transparency*, which requires up-to-date documentation and information sharing with the AI Office; (2) *copyright*, which requires legal use of content, compliance with rights, and complaint mechanisms; and (3) *safety and security*, which requires providers of high-risk GPAI models to assess and mitigate risks throughout the model lifecycle, implement safeguards, and report on their efforts. While these measures provide some safeguards, it is too early to assess whether they are sufficient. Comprehensive, concerted, collaborative, and multi-stakeholder (i.e., auditors, developers, providers, and policymakers) efforts, therefore, must be undertaken globally to ensure that these systems, like conventional AI, can be effectively and rigorously audited (Future of Life Institute, 2024).

---

[9] The AI Act, for example, defines GPAI as an *"AI model, including where such an AI model is trained with a large amount of data using self-supervision at scale, that displays significant generality and is capable of competently performing a wide range of distinct tasks regardless of the way the model is placed on the market and that can be integrated into a variety of downstream systems or applications"* (AI Act, 2024; § 3(63)). Here, the "generality" of a model is *"determined by a number of parameters, models with at least a billion of parameters and trained with a large amount of data using self-supervision at scale"* (AI Act, 2024; § 3(98)), and *"they allow for flexible generation of content, such as in the form of text, audio, images or video, that can readily accommodate a wide range of distinctive tasks"* (AI Act, 2024; § 3(99)).

Furthermore, many organizations that are in the process of developing and testing frontier AI systems—including GPAI models—are more likely to deploy these systems internally. Although internal deployment may be perceived as beneficial for these organizations, it can also pose significant risks, and Stix et al. (2025) argue that *"governance of the internal deployment of highly advanced frontier AI systems [is currently] absent."* The authors advocate the urgent implementation of robust governance policies that include a layered defense-in-depth approach, including scheming detection and control, internal use policies, oversight frameworks, targeted transparency, and disaster resilience plans to mitigate potential risks, including loss of control and societal disruption (Stix et al., 2025). Although several of these proposed measures fall under the purview of internal audits, similar to those mandated in existing AI regulations such as the EU AI Act, the challenge remains that internal audits are not independent, and their enforcement depends on the goodwill and willingness of the organization conducting them.

Finally, effective and fair AI auditability depends significantly on the independence, autonomy, accountability, and competence of the auditors responsible for auditing AI systems. Unfair practices, particularly those resulting from dual functions as auditors and consultants, represent a serious case of conflict of interest and may erode public trust in auditability and in AI. Consequently, several guidelines and best practices have been proposed to ensure that auditors are well equipped (through access to tools, methodologies, and competencies), have the necessary facilities (collaboration with internal auditors and access to model and data information) and safeguards to conduct fair and independent audits (Ojewale et al., 2024; Li and Goel, 2025). AI auditors are expected to have a diverse set of knowledge competencies and skills, ranging from technical knowledge of AI, statistics, data science, and software engineering, to knowledge of societal implications, business processes, and corporate governance, as well as expertise in regulatory and legal instruments (Li and Goel, 2025). In addition, effective enforcement of AI audits also requires access to standardized, context-specific, and holistic evaluation metrics and frameworks that can be translated into open-source, robust, and validated methodologies and toolkits that are not developed in an *ad hoc* manner, but integrated into an ecosystem of common and standardized audit infrastructure (Ojewale et al., 2024). Alongside the tools and infrastructure for auditing, auditors should be empowered by the regulatory frameworks to access proprietary information about the data, the model, and the model's interaction with the input data, as well as appropriate means to communicate their findings to a wide audience, including the general public (Ojewale et al., 2024). Reforms are also sought in international AI regulations that provide adequate safeguards and protections for auditors, audit tool developers, and rating agencies, including provisions for the organization of community action and appropriate accountability measures in response to detected harms or misconduct (Ojewale et al., 2024). The accomplishment of these guidelines and provisions requires a global collaboration among auditors, standards organizations, AI developers, and policymakers.

These are the fundamental changes that will need to be made, and the requirements that will need to be met, in order to facilitate a positive shift towards the effective operationalization of AI audits and auditability.

## Abbreviations

AI: Artificial Intelligence
ALTAI: Assessment List for Trustworthy AI
CAHAI: Ad Hoc Committee on Artificial Intelligence
CAI: Committee on Artificial Intelligence
CE: Conformité européenne
CEN: European Committee for Standardization
CENELEC: European Electrotechnical Committee for Standardization
CoE: Council of Europe
COMPAS: Correctional Offender Management Profiling for Alternative Sanctions
DPIA: Data Protection Impact Assessment
DUO: Dutch Executive Agency for Education
ESG: Environmental, Social, and Governance
EU: European Union
EC: European Commission
FRIA: Fundamental Rights Impact Assessment
GDPR: General Data Protection Regulation

GPAI: General Purpose Artificial Intelligence
HLEG AI: High-Level Expert Group on Artificial Intelligence
HREC: Human Research Ethics Committee
HRS: High-Risk AI Systems
IEC: International Electrotechnical Commission
IEEE: Institute of Electrical and Electronics Engineers
ISO: International Organization for Standardization
ITU: International Telecommunications Union
KPI: Key Performance Indicator
LLM: Large Language Model
NIST: National Institute of Standards and Technology
NB: Notified Body
OECD: Organization for Economic Cooperation and Development
SME: Small and Medium Enterprise
UN: United Nations

# References

Accessibility Act (2019). Directive (EU) 2019/882 of the European Parliament and of the Council of 17 April 2019 on the accessibility requirements for products and services (Accessibility Act). European Union.
https://eur-lex.europa.eu/legal-content/EN/TXT/?uri=CELEX%3A32019L0882

AI Act (2024). Regulation (EU) 2024/1689 of the European Parliament and of the Council of 13 June 2024 laying down harmonised rules on artificial intelligence and amending Regulations (EC) No 300/2008, (EU) No 167/2013, (EU) No 168/2013, (EU) 2018/858, (EU) 2018/1139 and (EU) 2019/2144 and Directives 2014/90/EU, (EU) 2016/797 and (EU) 2020/1828 (Artificial Intelligence Act). European Union.
https://eur-lex.europa.eu/legal-content/EN/TXT/?uri=CELEX%3A32024R1689

AI HLEG (2019, April 8). Ethics guidelines for trustworthy AI. High-Level Expert Group on AI.
https://digital-strategy.ec.europa.eu/en/library/ethics-guidelines-trustworthy-ai

AI HLEG (2020, July 17). Assessment List for Trustworthy Artificial Intelligence (ALTAI) for Self-Assessment. High-Level Expert Group on AI.
https://digital-strategy.ec.europa.eu/en/library/assessment-list-trustworthy-artificial-intelligence-altai-self-assessment

Algorithm Audit (2024). Preventing prejudice. Algorithm Audit.
https://algorithmaudit.eu/algoprudence/cases/aa202401_preventing-prejudice/

Arnold, M., Bellamy, R. K., Hind, M., Houde, S., Mehta, S., Mojsilović, A., Nair, R., Ramamurthy, K. Natesan, Olteanu, A., Piorkowski, D., Reimer, D., Richards, J., Tsay, J., & Varshney, K. R. (2019). FactSheets: Increasing trust in AI services through supplier's declarations of conformity. IBM Journal of Research and Development, 63(4/5), 6-1.

Australian Government (2024). Proposal Paper for Introducing Mandatory Guardrails for AI in High-Risk Setting. Australian Government Department of Industry, Science, and Resources.
https://consult.industry.gov.au/ai-mandatory-guardrails

Ayling, J., & Chapman, A. (2022). Putting AI ethics to work: are the tools fit for purpose?. AI and Ethics, 2(3), 405-429.

Balahur, A., Jenet, A., Hupont, I. T., Charisi, V., Ganesh, A., Griesinger, C. B., Maurer, P., Mian, L., Salvi, M., Scalzo, S., Josep, G. S., Fabio, T., & Tolan, S. (2022). Data quality requirements for inclusive, non-biased and trustworthy AI. HAL preprint hal-03889925.

Bender, E. M., & Friedman, B. (2018). Data statements for natural language processing: Toward mitigating system bias and enabling better science. Transactions of the Association for Computational Linguistics, 6, 587-604.

Bender, E. M., Gebru, T., McMillan-Major, A., & Shmitchell, S. (2021, March). On the dangers of stochastic parrots: Can language models be too big?🦜. In Proceedings of the 2021 ACM conference on fairness, accountability, and transparency (pp. 610-623).

Berghoff, C., Böddinghaus, J., Danos, V., Davelaar, G., Doms, T., Ehrich, H., Forrai, A., Grosu, R., Hamon, R., Junklewitz, H., Neu, M., Romanski, S., Samek, W., Schlesinger, D., Stavesand, J-E., Steinbach, S., von Twickel, A., Walter, R., Weissenböck, J., Wenzel, M., Wiegand, T. (2021). Towards Auditable AI Systems: From Principles to Practice [White Paper]. Fraunhofer Forum Digitale Technologien, Berlin, organized by the Federal Office for Information Security Germany, the TÜV-Verband and the Fraunhofer HHI. https://iphome.hhi.de/samek/pdf/BerAudit22.pdf

Brown, S., Davidovic, J., & Hasan, A. (2021). The algorithm audit: Scoring the algorithms that score us. Big Data & Society, 8(1), 2053951720983865.

Buolamwini, J., & Gebru, T. (2018, January). Gender shades: Intersectional accuracy disparities in commercial gender classification. In Conference on fairness, accountability and transparency (pp. 77-91). PMLR.

Burgess, M., Schot, E., & Geiger, G. (2023, March 06). This Algorithm Could Ruin Your Life. Wired. https://www.wired.com/story/welfare-algorithms-discrimination/

CAI (2024, November 28). Methodology for the Risk and Impact Assessment of Artificial Intelligence Systems from the Point of View of Human Rights, Democracy, and Rule of Law (HUDERIA Methodology). Committee on AI - Council of Europe. https://rm.coe.int/cai-2024-16rev2-methodology-for-the-risk-and-impact-assessment-of-arti/1680b2a09f

Cai, W., Encarnacion, R., Chern, B., Corbett-Davies, S., Bogen, M., Bergman, S., & Goel, S. (2022, June). Adaptive sampling strategies to construct equitable training datasets. In Proceedings of the 2022 ACM Conference on Fairness, Accountability, and Transparency (pp. 1467-1478).

Castets-Renard, C., & Besse, P. (2022). Ex Ante accountability of the AI Act: Between certification and standardization, in pursuit of fundamental rights in the country of compliance. Pursuit of Fundamental Rights in the Country of Compliance (August 29, 2022). Artificial Intelligence Law: Between Sectoral Rules and Comprehensive Regime. Comparative Law Perspectives, C. Castets-Renard & J. Eynard (eds), Bruylant, Forthcoming.

Catania, P., Lee, K., & Yu, A. (2024, October 08). Australia releases proposed mandatory guardrails for AI regulation. Corrs Chambers Westgarth. https://www.corrs.com.au/insights/australia-releases-proposed-mandatory-guardrails-for-ai-regulation

Chmielinski, K. S., Newman, S., Taylor, M., Joseph, J., Thomas, K., Yurkofsky, J., & Qiu, Y. C. (2022). The dataset nutrition label (2nd gen): Leveraging context to mitigate harms in artificial intelligence. arXiv preprint arXiv:2201.03954.

Civil Rights Act (1964). Title VI, Civil Rights Act of 1964. U.S. Department of Labor. https://www.dol.gov/agencies/oasam/regulatory/statutes/title-vi-civil-rights-act-of-1964

Cobbe, J., Lee, M. S. A., & Singh, J. (2021, March). Reviewable automated decision-making: A framework for accountable algorithmic systems. In Proceedings of the 2021 ACM conference on fairness, accountability, and transparency (pp. 598-609).

Constantinides, M., Bogucka, E., Quercia, D., Kallio, S., & Tahaei, M. (2024). RAI guidelines: Method for generating responsible AI guidelines grounded in regulations and usable by (non-) technical roles. Proceedings of the ACM on Human-Computer Interaction, 8(CSCW2), 1-28.

Cooper, A. F., Moss, E., Laufer, B., & Nissenbaum, H. (2022, June). Accountability in an algorithmic society: relationality, responsibility, and robustness in machine learning. In Proceedings of the 2022 ACM conference on fairness, accountability, and transparency (pp. 864-876).

Council of Europe (2024a, September 5). The Framework Convention on Artificial Intelligence and Human Rights, Democracy and the Rule of Law. Council of Europe. https://rm.coe.int/1680afae3c and https://www.coe.int/en/web/artificial-intelligence/the-framework-convention-on-artificial-intelligence

Council of Europe (2024b, December 2). HUDERIA: New tool to assess the impact of AI systems on human rights. Council of Europe. https://www.coe.int/en/web/portal/-/huderia-new-tool-to-assess-the-impact-of-ai-systems-on-human-rights

Cybersecurity and Resilience Bill (2024). Cybersecurity and Resilience Bill. UK Department for Science, Innovation and Technology. https://www.gov.uk/government/collections/cyber-security-and-resilience-bill

Data Protection Act (2018). UK Data Protection Act. https://www.legislation.gov.uk/ukpga/2018/12/contents/enacted

Delcker, J. (2025, January 28). AI: US under Trump and Europe choose diverging paths. DW. https://www.dw.com/en/ai-policy-regulation-europe-us/a-71426911

Digital Services Act (2022). Regulation (EU) 2022/2065 of the European Parliament and of the Council of 19 October 2022 on a Single Market For Digital Services and amending Directive 2000/31/EC (Digital Services Act). European Union. https://eur-lex.europa.eu/eli/reg/2022/2065/oj/eng

Eulerich, M. (2021). The new three lines model for structuring corporate governance–A critical discussion of similarities and differences. Available at SSRN 3777392.

Doshi-Velez, F., & Kim, B. (2018). Considerations for evaluation and generalization in interpretable machine learning. Explainable and interpretable models in computer vision and machine learning, 3-17.

Dressel, J., & Farid, H. (2018). The accuracy, fairness, and limits of predicting recidivism. Science advances, 4(1), eaao5580.

Dwivedi, R., Dave, D., Naik, H., Singhal, S., Omer, R., Patel, P., Qian, B., Wen, Z., Shah, T., Morgan, G., & Ranjan, R. (2023). Explainable AI (XAI): Core ideas, techniques, and solutions. ACM Computing Surveys, 55(9), 1-33.

Ellul, J., Pace, G., McCarthy, S., Sammut, T., Brockdorff, J., & Scerri, M. (2021, June). Regulating artificial intelligence: a technology regulator's perspective. In Proceedings of the eighteenth international conference on artificial intelligence and law (pp. 190-194).

Employment Equality Directive (2000). Council Directive 2000/78/EC of 27 November 2000 establishing a general framework for equal treatment in employment and occupation. European Union. https://eur-lex.europa.eu/eli/dir/2000/78/oj/eng

European Commission (2024, August 1). AI Act. https://digital-strategy.ec.europa.eu/en/policies/regulatory-framework-ai

Field, H. (2021, September 27). Behind the scenes: How Twitter decided to open up its image-cropping algorithm to the public. Tech Brew. https://www.emergingtechbrew.com/stories/2021/09/27/behind-the-scenes-of-twitter-s-decision-to-open-up-its-image-cropping-algorithm-to-researchers

Floridi, L., Holweg, M., Taddeo, M., Amaya, J., Mökander, J., & Wen, Y. (2022). CapAI-A procedure for conducting conformity assessment of AI systems in line with the EU artificial intelligence act. Available at SSRN 4064091.

FRA (2009). EU Charter of Fundamental Rights. European Union Agency for Fundamental Rights. https://fra.europa.eu/en/eu-charter

Future of Life Institute (2023, March 22). Pause Giant AI Experiments: An Open Letter. https://futureoflife.org/open-letter/pause-giant-ai-experiments/

Future of Life Institute (2024). EU Artificial Intelligence Act - An introduction to Codes of Practice for the AI Act. https://artificialintelligenceact.eu/introduction-to-codes-of-practice/

Future of Life Institute (2025). EU Artificial Intelligence Act - Standard Setting. https://artificialintelligenceact.eu/standard-setting/

Gabriel, I., & Keeling, G. (2025). A matter of principle? AI alignment as the fair treatment of claims. Philosophical Studies, 1-23.

GDPR (2016). Regulation (EU) 2016/679 of the European Parliament and of the Council of 27 April 2016 on the protection of natural persons with regard to the processing of personal data and on the free movement of such data, and repealing Directive 95/46/EC (General Data Protection Regulation). European Union. https://eur-lex.europa.eu/eli/reg/2016/679/oj/eng

Gebru, T., Morgenstern, J., Vecchione, B., Vaughan, J. W., Wallach, H., Iii, H. D., & Crawford, K. (2021). Datasheets for datasets. Communications of the ACM, 64(12), 86-92.

Gheibi, O., & Weyns, D. (2024). Dealing with drift of adaptation spaces in learning-based self-adaptive systems using lifelong self-adaptation. ACM Transactions on Autonomous and Adaptive Systems, 19(1), 1-57.

Heikkilä, M. (2022, March 29). Dutch scandal serves as a warning for Europe over risks of using algorithms. POLITICO. https://www.politico.eu/article/dutch-scandal-serves-as-a-warning-for-europe-over-risks-of-using-algorithms/

Herrera-Poyatos, A., Del Ser, J., de Prado, M. L., Wang, F. Y., Herrera-Viedma, E., & Herrera, F. (2025). Responsible Artificial Intelligence Systems: A Roadmap to Society's Trust through Trustworthy AI, Auditability, Accountability, and Governance. arXiv preprint arXiv:2503.04739.

Holland, S., Hosny, A., Newman, S., Joseph, J., & Chmielinski, K. (2020). The dataset nutrition label. Data protection and privacy, 12(12), 1.

Holstege, F., Jorgensen, M., Padh, K., Parie, J., Persson, J., Prorokovic, K., & Snoek, L. (2025). Auditing a Dutch Public Sector Risk Profiling Algorithm Using an Unsupervised Bias Detection Tool. arXiv preprint arXiv:2502.01713.

Hutchinson, B., Smart, A., Hanna, A., Denton, E., Greer, C., Kjartansson, O., Barnes, P., & Mitchell, M. (2021, March). Towards accountability for machine learning datasets: Practices from software engineering and infrastructure. In Proceedings of the 2021 ACM conference on fairness, accountability, and transparency (pp. 560-575).

IEEE SA (2021). IEEE 7000: IEEE Standard Model Process for Addressing Ethical Concerns during System Design. IEEE Standards Association. https://standards.ieee.org/ieee/7000/6781/

International Organization for Standardization (2023). ISO/IEC 42001: 2023 - Information Technology - Artificial Intelligence - Management System. https://www.iso.org/standard/81230.html

ITU (2024). AI Governance Day - From Principles to Implementation. International Telecommunications Union - Telecommunication Standardization Sector. Retrieved from: https://s41721.pcdn.co/wp-content/uploads/2021/06/2401225_AI_Governance_Day_2024_Report-E.pdf

Jaiswal, S., Duggirala, K., Dash, A., & Mukherjee, A. (2022, May). Two-face: Adversarial audit of commercial face recognition systems. In Proceedings of the International AAAI Conference on Web and Social Media (Vol. 16, pp. 381-392).

Kilian, R., Jäck, L., & Ebel, D. (2025). European AI Standards-Technical Standardization and Implementation Challenges under the EU AI Act. Available at SSRN 5155591.

Knowles, B., & Richards, J. T. (2021, March). The sanction of authority: Promoting public trust in AI. In Proceedings of the 2021 ACM conference on fairness, accountability, and transparency (pp. 262-271).

Königstorfer, F., & Thalmann, S. (2022). AI Documentation: A path to accountability. Journal of Responsible Technology, 11, 100043.

LaBrie, R. C., & Steinke, G. (2019). Towards a framework for ethical audits of AI algorithms. In Proceedings of 25th Americas Conference on Information Systems (pp. 1-5).

Lau, T. (2020, April 01). Predictive Policing Explained. Brennan Center for Justice. https://www.brennancenter.org/our-work/research-reports/predictive-policing-explained

Larson J., Mattu, S., Kirchner, L., & Angwin, J. (2016, May 23). How we Analyzed the COMPAS Recidivism Algorithm. ProPublica. https://www.propublica.org/article/how-we-analyzed-the-compas-recidivism-algorithm

Leslie, D., Burr, C., Aitken, M., Cowls, J., Katell, M., & Briggs, M. (2021). Artificial intelligence, human rights, democracy, and the rule of law: a primer. arXiv preprint arXiv:2104.04147.

Li, B., Qi, P., Liu, B., Di, S., Liu, J., Pei, J., Yi, J., & Zhou, B. (2023). Trustworthy AI: From principles to practices. ACM Computing Surveys, 55(9), 1-46.

Li, Y. (2017). Deep reinforcement learning: An overview. arXiv preprint arXiv:1701.07274.

Li, Y., & Goel, S. (2024). Making it possible for the auditing of ai: A systematic review of ai audits and ai auditability. *Information Systems Frontiers*, 1-31.

Li, Y., & Goel, S. (2025). Artificial intelligence auditability and auditor readiness for auditing artificial intelligence systems. International Journal of Accounting Information Systems, 56, 100739.

Liu, H., Wang, Y., Fan, W., Liu, X., Li, Y., Jain, S., Liu, Y., Jain, A., & Tang, J. (2022). Trustworthy ai: A computational perspective. ACM Transactions on Intelligent Systems and Technology, 14(1), 1-59.

Luccioni, A. S., Pistilli, G., Sefala, R., & Moorosi, N. (2025). Bridging the Gap: Integrating Ethics and Environmental Sustainability in AI Research and Practice. arXiv preprint arXiv:2504.00797.

Macdonald, A. (2025, May 17). Advocates pick privacy, inclusion holds in Kenya's Maisha Namba digital ID system. biometricupdate.com. https://www.biometricupdate.com/202503/advocates-pick-privacy-inclusion-holds-in-kenyas-maisha-namba-digital-id-system

Migliorini, S. (2024). China's Interim Measures on generative AI: Origin, content and significance. Computer Law & Security Review, 53, 105985.

Minkkinen, M., Niukkanen, A., & Mäntymäki, M. (2024). What about investors? ESG analyses as tools for ethics-based AI auditing. *AI & society, 39*(1), 329-343.

Mitchell, M., Wu, S., Zaldivar, A., Barnes, P., Vasserman, L., Hutchinson, B., Spitzer, E., Raji, I. D., & Gebru, T. (2019, January). Model cards for model reporting. In Proceedings of the conference on fairness, accountability, and transparency (pp. 220-229).

Mökander, J. (2023). Auditing of AI: Legal, ethical and technical approaches. *Digital Society*, *2*(3), 49.

Mökander, J., & Floridi, L. (2021). Ethics-based auditing to develop trustworthy AI. Minds and Machines, 31(2), 323-327.

Mökander, J., & Floridi, L. (2023). Operationalising AI governance through ethics-based auditing: an industry case study. *AI and Ethics, 3*(2), 451-468.

Mökander, J., Schuett, J., Kirk, H. R., & Floridi, L. (2024). Auditing large language models: a three-layered approach. AI and Ethics, 4(4), 1085-1115.

Munoko, I., Brown-Liburd, H. L., & Vasarhelyi, M. (2020). The ethical implications of using artificial intelligence in auditing. *Journal of business ethics*, *167*(2), 209-234.

Naundorf, K. (2023). The Twisted Eye in the Sky Over Buenos Aires. *Wired UK*, *13*. https://www.wired.com/story/buenos-aires-facial-recognition-scandal/

NIST (2023, January 26). AI Risk Management Framework. National Institute of Standards and Technology. https://www.nist.gov/itl/ai-risk-management-framework

NL Times (2024, November 12). Government to refund over 10,000 students over discriminatory DUO fraud checks. NL Times. https://nltimes.nl/2024/11/12/government-refund-10000-students-discriminatory-duo-fraud-checks

OECD (2019). OECD AI Principles for Trustworthy AI. Organization for Economic Cooperation and Development AI Policy Observatory. https://oecd.ai/en/ai-principles

OECD (2024). Recommendation of the Council on Artificial Intelligence. Organization for Economic Co-Operation and Development. https://legalinstruments.oecd.org/en/instruments/OECD-LEGAL-0449

Ojewale, V., Steed, R., Vecchione, B., Birhane, A., & Raji, I. D. (2024). Towards AI accountability infrastructure: Gaps and opportunities in AI audit tooling. *arXiv preprint arXiv:2402.17861*.

Palmer, S. (2025). Grok 3: The Case for an Unfiltered AI Model. https://shellypalmer.com/2025/03/grok-3-the-case-for-an-unfiltered-ai-model/

Pavlidis, G. (2024). Unlocking the black box: analysing the EU artificial intelligence act's framework for explainability in AI. Law, Innovation and Technology, 16(1), 293-308.

Pégny, M., Thelisson, E., & Ibnouhsein, I. (2019). The Right to an Explanation: An Interpretation and Defense. Delphi, 2, 161.

Pentland, B. T. (1993). Getting comfortable with the numbers: Auditing and the micro-production of macro-order. Accounting, organizations and society, 18(7-8), 605-620.

Petkovic, D. (2023). It is not "Accuracy vs. Explainability"—we need both for trustworthy AI systems. IEEE Transactions on Technology and Society, 4(1), 46-53.

Power, M. (1996). Making things auditable. Accounting, organizations and society, 21(2-3), 289-315.

Power, M. K. (2003). Auditing and the production of legitimacy. Accounting, organizations and society, 28(4), 379-394.

ProPublica (2016, May 23). Machine Bias – There's software used across the country to predict future criminals. And it's biased against blacks. ProPublica. https://www.propublica.org/article/machine-bias-risk-assessments-in-criminal-sentencing

Raji, I. D., Smart, A., White, R. N., Mitchell, M., Gebru, T., Hutchinson, B., Smith-Loud, J., Theron, D., & Barnes, P. (2020, January). Closing the AI accountability gap: Defining an end-to-end framework for internal algorithmic auditing. In Proceedings of the 2020 conference on fairness, accountability, and transparency (pp. 33-44).

Raji, I. D., Kumar, I. E., Horowitz, A., & Selbst, A. (2022, June). The fallacy of AI functionality. In Proceedings of the 2022 ACM Conference on Fairness, Accountability, and Transparency (pp. 959-972).

Raji, I. D., Xu, P., Honigsberg, C., & Ho, D. (2022, July). Outsider oversight: Designing a third party audit ecosystem for ai governance. In Proceedings of the 2022 AAAI/ACM Conference on AI, Ethics, and Society (pp. 557-571).

Scanell, B., Moore, L., and Hayes, R. (2024, July 18). The Time to (AI) Act is Now: A Practical Guide to Biometric Categorisation System Under the AI Act. William Fry. https://www.williamfry.com/knowledge/the-time-to-ai-act-is-now-a-practical-guide-to-biometric-categorisation-systems-under-the-ai-act/

Schaart, E. (2021, January 15). Mark Rutte pulls plug on Dutch government, plans immediate return. POLITICO. https://www.politico.eu/article/dutch-government-resigns-over-childcare-benefit-scandal/

Shneiderman, B. (2020). Bridging the gap between ethics and practice: guidelines for reliable, safe, and trustworthy human-centered AI systems. ACM Transactions on Interactive Intelligent Systems (TiiS), 10(4), 1-31.

Sloane, M., Moss, E., & Chowdhury, R. (2022). A Silicon Valley love triangle: Hiring algorithms, pseudo-science, and the quest for auditability. Patterns, 3(2).

Sloane, M., & Zakrzewski, J. (2022, June). German ai start-ups and "ai ethics": Using a social practice lens for assessing and implementing socio-technical innovation. In Proceedings of the 2022 ACM Conference on Fairness, Accountability, and Transparency (pp. 935-947).

Soler Garrido, J., De Nigris, S., Bassani, E., Sanchez, I., Evas, T., André, A-A., & Boulangé, T. (2024). Harmonised Standards for the European AI Act, European Commission. https://publications.jrc.ec.europa.eu/repository/handle/JRC139430

Stix, C., Pistillo, M., Sastry, G., Hobbhahn, M., Ortega, A., Balesni, M., Hallensleben, A., Goldowsky-Dil, N., & Sharkey, L. (2025). AI Behind Closed Doors: a Primer on The Governance of Internal Deployment. arXiv preprint arXiv:2504.12170.

Supreme Auditing Institutions (2023). Auditing Machine Learning Algorithms - A White Paper for Public Auditors. Supreme Audit Institutions of Finland, Germany, the Netherlands, Norway and the UK. https://www.auditingalgorithms.net/

TerraBrasilis – Geographic Data platform. (n.d.). https://terrabrasilis.dpi.inpe.br/

Thelisson, E. (2017, August). Towards Trust, Transparency and Liability in AI/AS systems. In IJCAI (pp. 5215-5216).

Thelisson, E. (2019). The scope of the extraterritorial character of the General Data Protection Regulation. Revue internationale de droit économique, (4), 501-533.

Thelisson, E. (2023). Non-prolifération de l'intelligence artificielle générative–Quel en-cadrement normatif?. In L'Association française pour l'Intelligence Artificielle, 120, 62-69.

Thelisson, E., Mika, G., Schneiter, Q., Padh, K., & Verma, H. (2023). Toward responsible AI use: Considerations for sustainability impact assessment. arXiv preprint arXiv:2312.11996.

Thelisson, E., & Verma, H. (2024). Conformity assessment under the EU AI act general approach. AI and Ethics, 4(1), 113-121.

UN (2024, May 03). United Nations system white paper on artificial intelligence governance: an analysis of current institutional models and related functions and existing international normative frameworks within the United Nations system that are applicable to artificial intelligence governance. United Nation Systems - Chief Executives Board for Coordination. https://unsceb.org/sites/default/files/2024-11/UNSystemWhitePaperAIGovernance.pdf

US Congress (2023). S.2892 - Algorithmic Accountability Act of 2023. Senate Committee on Commerce, Science, and Transportation. https://www.congress.gov/bill/118th-congress/senate-bill/2892/text

Vincent, J. (2018, June 27). IBM hopes to fight bias in facial recognition with new diverse dataset. The Verge. https://www.theverge.com/2018/6/27/17509400/facial-recognition-bias-ibm-data-training

Washington, A. L. (2018). How to argue with an algorithm: Lessons from the COMPAS-ProPublica debate. Colo. Tech. LJ, 17, 131.

White & Case (2024, December 16). AI Watch: Global regulatory tracker - Australia. White & Chase. https://www.whitecase.com/insight-our-thinking/ai-watch-global-regulatory-tracker-australia

White & Case (2025, May 29). AI Watch: Global regulatory tracker - China. White & Case. https://www.whitecase.com/insight-our-thinking/ai-watch-global-regulatory-tracker-china

White House (2023, November 01). Safe, Secure, and Trustworthy Development and Use of Artificial Intelligence (Executive Order 14110 of October 30, 2023). National Archives - Federal Register: Presidential Document. https://www.federalregister.gov/documents/2023/11/01/2023-24283/safe-secure-and-trustworthy-development-and-use-of-artificial-intelligence

White House (2025, January 23). Removing Barriers to American Leadership in Artificial Intelligence. President Donald J. Trump Executive Order. https://www.whitehouse.gov/presidential-actions/2025/01/removing-barriers-to-american-leadership-in-artificial-intelligence/

Wolf, M. J., Miller, K., & Grodzinsky, F. S. (2017). Why we should have seen that coming: comments on Microsoft's tay" experiment," and wider implications. Acm Sigcas Computers and Society, 47(3), 54-64.

Wong, R. Y., Madaio, M. A., & Merrill, N. (2023). Seeing like a toolkit: How toolkits envision the work of AI ethics. Proceedings of the ACM on Human-Computer Interaction, 7(CSCW1), 1-27.

Yurrita, M., Murray-Rust, D., Balayn, A., & Bozzon, A. (2022, June). Towards a multi-stakeholder value-based assessment framework for algorithmic systems. In Proceedings of the 2022 ACM Conference on Fairness, Accountability, and Transparency (pp. 535-563).